\documentclass{svproc}
\usepackage{xcolor}
\usepackage{amsmath,amssymb, bbm}
\usepackage{epsfig}
\usepackage{amssymb}
\usepackage{graphicx,float}
\usepackage{subcaption}
\usepackage{url}

\date{}
\begin{document}

\mainmatter

\title{Network Embedding Analysis for Anti-Money Laundering Detection}

\author{Anthony Bonato\inst{1}\thanks{Supported by an NSERC Discovery Grant.} \and Adam Szava\inst{1}}

\institute{Toronto Metropolitan University, Toronto, Ontario, Canada}

\maketitle

\begin{abstract}
We employ network embedding to detect money laundering in financial transaction networks. Using real anonymized banking data, we model over one million accounts as a directed graph and use it to refine previously detected suspicious cycles with node2vec embeddings, creating a new network parameter, the spread number. Combined with more traditional centrality measures, these define an aggregate score $R$ that highlights so-called anti-central nodes: accounts that are structurally important yet organized to avoid detection. Our results show only a small subset of cycles attain high $R$ values, flagging concentrated groups of suspicious accounts. Our approach demonstrates the potential of embedding-based network analysis to expose laundering strategies that evade traditional graph centrality measures.
\end{abstract}

\section{Introduction}\label{intro}

The concealment of the origin of illegally obtained funds used in the legal financial system, commonly known as money laundering, is a global phenomenon. While it is evident that this illegal scheming could negatively impact the domestic economy where it occurs, the exact scale and nature of its effect are hard to measure. The results in \cite{Beqiri2018} highlight various economic factors that could be affected by money laundering. Laundered money is not properly taxed, resulting in impaired budget inflows for the local government. Money launderers are also often not interested in generating profits from their laundering investments, and rather are focused on protecting their earlier (illegal) cash-inflows. This behavior undermines productive investment and distorts market efficiency. Finally, money laundering has a negative impact on the local privatization process, as money launderers outbid and replace honest buyers, which in turn has a detrimental effect on the country's economy through mismanagement and economic inefficiencies.

The first effort to enforce financial regulation to prevent money laundering came with the 1970 Bank Secrecy Act \cite{white}. Since then, governmental agencies have been established to oversee \emph{anti-money laundering} (or AML) mandates, either individually by national governments or through intergovernmental institutions such as the Financial Action Task Force (or FATF) \cite{murrar}. These regulatory institutions rely on self-reported data provided by the banks to enforce the inspection of compliance frameworks, risk management policies, and transaction monitoring of high-risk or illicit activity \cite{nojeim,schneider}. The banks must then employ research and tools to identify which transactions to report, or face sanctions from these regulators. This is important and relevant to modern-day banks as evidenced by the 2024 settlement, which ordered TD Bank to pay more than $\$3$ billion for failing to adequately guard against money laundering between the years 2014-2023, allowing for more than $\$670$ million of illicit funds to flow through their banking system \cite{tdbank}.

Tools from network science, such as network embeddings and centrality measures, offer an opportunity to algorithmically assist in detecting potential money laundering transactions. Such methods do not rely on a large set of ground-truth labeled data, which is often private and therefore not readily available \cite{jullum}.

Rabobank was formed in 1972 from the merger of the Coöperatieve Centrale Raiffeisen-Bank (Utrecht) and the Coöperatieve Centrale Boerenleenbank (Eindhoven), federations comprising over a thousand local cooperative banks \cite{degraaf}. We use an anonymized Rabobank banking transaction data set provided by \cite{saxena}. The data set consists of transactions between accounts grouped by directed account pairs, with each transaction having the following data:
\begin{enumerate}
    \item Sender and Receiver Account IDs: A pair of account IDs, which can be denoted by a directed edge $(u,v)$ with a start node $u$ and an end node $v$.
    \item Number of Transactions: A positive integer $n$, the number of transactions from sender to receiver.
    \item Amount of Money Transferred: A positive edge weight $k$, corresponding to the total amount of money in undisclosed \$ currency,  transferred from the start node to the end node from 2010 to 2020.
    \item Start \& End Years: The years corresponding to the first and last transaction for each directed edge.
\end{enumerate}

In the goal of identifying new suspicious transactions and their associated nodes, we focus on the transactions that are not already being reported to the regulatory agencies. To this end, we consider only account pairs with an average amount of money transferred less than \$10,000, and all transactions between said accounts taking place in the same year, where the start year equals the end year. See previous work in \cite{prev} for further discussion on how the data was prepared. The final cleaned data set forms a graph with $1,048,166$ accounts and $1,686,184$ directed edges.

Previous work in \cite{prev} identified a relatively small set of transaction cycles in the Rabobank network whose constituent nodes were flagged as suspicious. One goal of this paper is to narrow down this set of suspicious nodes using more advanced tools from network science. Specifically, we aim to gain further insight into the structural role of suspicious cycles within their local communities by applying network embedding techniques. Unlike previous work, this approach aims to identify nodes whose structural importance becomes evident in the embedding space, even when they are otherwise organized to avoid seeming central according to traditional measures.

The paper is organized as follows. In Section~2, we will review the principles of AML detection based on the current working model of money laundering strategies. Additionally, we will review our previous work, as presented in \cite{prev}, and a review of the related literature. Section~3 will present new techniques using graph embeddings and centrality measures, followed by Section~4, which will present the results of applying these techniques to the Rabobank network. Finally, in Section~5, we will discuss the results and consider future directions.

We consider finite directed graphs (or digraphs) in this paper. Additional background on graph theory can be found in \cite{west}, and further background on complex networks is available in the book \cite{bonato}.

\section{Literature Review}\label{aml}

AML is concerned with detecting accounts associated with money laundering within a banking system. To be concrete, the banking industry models such accounts as those that take part in a three-step process that the illicit money will follow to try to infiltrate the legal banking network \cite{cassella,schneider}.
\begin{enumerate}
    \item \textit{Placement:} The money launderer obtains illegal profit, and inserts the cash into the financial system (possibly through cash-intensive businesses like restaurants or gas stations).
    \item \textit{Layering:} The money then moves through a series of transactions to obfuscate the trail of money, making it harder to trace if some institution were to attempt to trace it back.
    \item \textit{Integration:} The money is then used for legitimate transactions to pay for goods and services.
\end{enumerate}
The work presented in this paper (and previous work done in \cite{prev}) focuses on detecting transactions that occur as part of the \textit{layering} step of the model.

To motivate some of the methodology that will be described in this paper, we also consider some other data-driven characterizations of money laundering behaviors in a transaction network. We would expect money launderers to avoid connections with high-degree accounts, and not restrict the number of transactions or geography of said transactions; see \cite{granados22}. Further, such accounts would exhibit unusual transaction behavior in terms of complexity, value, and deviation from known customer patterns; see \cite{ogbeide24}.

While the broad topic of AML has been extensively studied, the role of network science within this field has been a relatively recent development. Work done by Colladon and Remondi \cite{colladon} represents an early attempt. They applied social network analysis to 19 months of transaction data between companies, showing that graph metrics such as degree, closeness, betweenness, and network constraint reveal high-risk clients more effectively than transaction attributes alone.

The $2019$ study presented in \cite{elliptic} introduced an important dataset of Bitcoin transactions (the \textit{Elliptic Data Set}) labeled with licit and illicit transactions. The authors demonstrated the effectiveness of \textit{Graph Convolutional Networks} for classifying transactions based on node features and graph structure.

Recent work in \cite{capozzi} describes \textit{WeirdFlows}, a search pipeline for detecting fraudulent transactions in a banking network, without the need for a training set. The pipeline generates a network from the transaction data and then generates multiple temporal networks by aggregating transfers over time, with the goal of discovering all the transaction paths of fixed length between nodes and analyzing them as a timeseries. This leads to interpretable results for AML analysts. The method was shown to be effective using real data.

In our previous work in \cite{prev}, we looked at the same Rabobank dataset by detecting communities within the large network and then within each major community (of which there were $11,827$), detecting directed cycles of transactions (of which there were $83$, varying in length from $3$ to $6$). This resulted in the flagging of $155$ accounts as being suspicious.

\section{Methods}\label{methods}

Networks may possess an explicit geometric structure; for instance, a road network can be drawn on a map with nodes positioned according to their relative positions in space. In general, network embedding is concerned with assigning a geometric representation to any arbitrary graph, typically in a low-dimensional Euclidean space, in a way that emphasizes important features.

More formally, a network embedding for a given (potentially weighted and directed) graph $G = (V(G), E(G))$ is a function $\mathcal{E}: V(G) \to \mathbb{R}^d$ for some fixed $d$ (which is the dimension of our embedding, and is typically less than the order of the graph). The assumption is that a low-dimensional space, referred to as the \textit{latent space}, exists where node information, such as centrality, community, and local structure, can be optimally encoded into geometric information. Embedding algorithms define $\mathcal{E}$ in various ways to approximate the latent space with what are known as \textit{feature spaces}.

A central problem is extracting the desired information from the graph. One popular technique is that of \textit{random walks}, where a large number of fixed-length random walks are generated starting from each node, from which node relationships can then be inferred. Nodes that frequently appear in each other's random walk collections are likely to be embedded by $\mathcal{E}$ into similar feature representations. Random walks in the context of embeddings were initially developed in \cite{deepwalk} for \textit{DeepWalk}; here, we use a different implementation of this technique, which was presented in \cite{node2vec} and is called \textit{node2vec}.

The algorithm \textit{node2vec} is formulated as a maximum likelihood optimization problem to generate the embedding $\mathcal{E} : V \to \mathbb{R}^d$, which can be encoded as a $|V|\times d$ matrix.
\begin{enumerate}
    \item Begin by iterating over each node $v \in V$ and generate its neighborhood $N_S(v)$ using a sampling strategy $S$ (sampling strategies can be tuned with two parameters $p,q$, discussed next).
    \item Use stochastic gradient ascent over the embedding parameters (the entries of the $|V|\times d$ matrix) to optimize:
    $$\max_\mathcal{E} \sum_{u\in V}\log\mathbb{P}\left(N_S(u)|\mathcal{E}(u)\right),$$
    where $\mathbb{P}\left(N_S(u)|\mathcal{E}(u)\right)$ is the probability of observing a network neighborhood $N_S(u)$ for a node $u$, conditioned on its feature representation $\mathcal{E}(u)$.
\end{enumerate}

An important consideration when using \textit{node2vec} is the random sampling technique $S$, which is parameterized by $p$ and $q$. The sampling technique extracts graph information, such as centrality and community, and is designed in \textit{node2vec} to be tunable, emphasizing certain properties. At the two extremes, we have BFS and DFS as sampling strategies, which are suitable for structural equivalence and homophily, respectively. The sampling strategy for \textit{node2vec} enables us to interpolate between these two paradigms, capturing both aspects of network information.

At each stage of the random walk, all outgoing edges from the current node are assigned a transition probability. Let $w_{ij}$ denote the weight of the (potentially directed) edge $(i,j) \in E(G)$, and let $d_{ij}$ denote the shortest path distance between nodes $i,j \in V(G)$.  Consider a random walk that just traversed edge $(t,v)$ and is now at node $v$, then for each $x\in V(G)$ such that $(v,x)\in E(G)$ assign outgoing transition probabilities by:
$$\mathbb{P}(x \text{ is next in the random walk}) = \frac{w_{vx}}{Z} \times\begin{cases}
    1/p & \text{if }d_{tx} = 0, \\
    1 & \text{if }d_{tx} = 1, \\
    1/q & \text{if }d_{tx} = 2, \\
\end{cases}$$
where $Z$ is a normalizing constant. That being said, we can intuitively describe the effect that the parameters $p$ and $q$ have on the random walk (and thus the embedding).
\begin{enumerate}
    \item $p$ (\emph{return parameter}): This parameter controls the likelihood of immediately revisiting a node in the walk; keeping this value low ($p < \min(q,1)$) would keep the walk ``local.''
    \item $q$ (\emph{in-out parameter}): This parameter allows us to interpolate between BFS ($q > 1$) and DFS ($q < 1$) strategies.
\end{enumerate}
There are many other network embedding algorithms which rely on various other techniques: matrix factorization methods are used in algorithms like \textit{Laplacian Eigenmaps} and \textit{GraRep}, random walks are used in other algorithms like \textit{Walklets}, and deep learning techniques were also used in \textit{DiffPool} and \textit{PALE}. See \cite{baptista2023} for a detailed review of current algorithms (including all the ones just mentioned). In this work, we choose to use \textit{node2vec}.

We next present a novel network parameter that incorporates information from the embedding space. We expect that money launderers would attempt to avoid detection. In particular, money launderers may strive not to stand out using typical centrality measures of the network (such as closeness, betweenness, and degree), while still representing influential nodes due to their structural position in the network. Such nodes would therefore be undetectable using traditional centrality measures, whose purpose was to identify key actors in the network. We refer to such nodes as \textit{anti-central}, and they must actively self-organize to avoid detection by typical centrality measures.

A useful concept is that of \textit{community normalization}. Given a centrality measure $\rho_K$ that assigns each node in a subgraph $K$ a value, we aim to extend this measure to a normalized centrality measure for the entire subgraph $H$. Let $G$ be a graph that has been partitioned into communities $\mathcal{P}$, and let $H$ be a subgraph entirely contained in some community. We define the \textit{community-normalized score} $\mathrm{CNS}(H, \rho_K)$ by the following process:
\begin{enumerate}
    \item Induce the subgraph of the community which $H$ belongs to, call this $P$.
    \item Compute $$r_{\mathrm{subgraph}}=\sum_{v \in H}\rho_P(v).$$
    \item Randomly sample $m$ subgraphs (for some fixed $m$) from $P-H$ of order $|V(H)|$, call this collection $\mathcal{S} = \{S_1, \dots, S_m\}$, and compute
    $$r_{\mathrm{random}} = \frac{1}{m}\sum_{i = 1}^m \left(\sum_{v \in S_i}\rho_P(v) \right).$$
    \item Let
    $$\mathrm{CNS}(H, \rho_K) = \frac{r_{\mathrm{subgraph}}}{r_{\mathrm{random}}}.$$
\end{enumerate}
Intuitively, a \textit{community-normalized score} ($\mathrm{CNS}$) greater than $1$ indicates that the subgraph $H$ scores higher according to $\rho_K$ compared to a random subgraph of the same size from within its community. In practice, it is possible for communities to be small, and so $P-H$ may have fewer nodes than $|V(H)|$. In such a case, just disregard this subgraph $H$.

Let $G$ be a weighted and directed transaction network, with a collection of cycles $\mathcal{C}$, and a partition into communities $\mathcal{P}$. For each cycle $C \in \mathcal{C}$, we compute an associated $r'$-value by $r' = \mathrm{CNS}(C, E)$ where $E$ is computed by
$$E(v)=\sum_{u \in V(C)} \sqrt{\sum_{k = 1}^d\left(\mathcal{E}(u)_k - \mathcal{E}(v)_k\right)^2},$$
where $\mathcal{E}$ is the embedding function. The value $E(v)$ computes the sum of the pairwise $\ell_2$ distances in the embedding space between $v$ and each other node in the cycle.
If $r' > 1$ for a given cycle, then the nodes in the cycle are more ``spread" apart in the feature space compared to a random collection of nodes from its community. At the end of this stage, each cycle $C \in \mathcal{C}$ has an associated $r'$ value.

Recall that the embedding function $\mathcal{E}$ generated by \textit{node2vec} depends on parameters $p,q$. Instead of fixing a specific choice for $p,q$, we opt to repeat the experiment $k$ times by varying $p,q \in (0,2]^2$ for $k$ different values. Let $r$ denote the proportion of the $k$ experiments where $C$ was an outlier in terms of a high $r'$ value (say, in the $75$th percentile). We call $r$ the \textit{spread number} of a cycle $C$, and we use the notation $r(C).$ It could be the case that $r = 0$, indicating the corresponding cycle is never (after the $k$ trials) notably more spread compared to its community.

Anti-centrality focuses on nodes that appear innocuous when compared to classical measures of centrality. To this end, we incorporate four traditional measures of centrality into our methods. For the following, consider a graph $G = (V,E)$ and let $v \in V$. Betweenness centrality considers nodes that are often found between two other nodes (along their shortest path between them) to be central. The \emph{betweenness} of $v$ is defined by
$$B(v)=\sum_{s,t \in V} \frac{\text{number of }s,t\text{ shortest paths that pass through }v}{\text{number of }s,t\text{ shortest paths}}.$$

\emph{Degree centrality} is defined by $D(v)= \frac{\text{deg}_G(v)}{|V|}.$
\emph{Closeness} considers nodes that are roughly close to all other nodes (via graph distance) to be central, and is defined by
$$\mathrm{CL}(v) = \frac{1}{\sum_{u \in V}d(u,v)}.$$

The \emph{common-out-neighbor} (or $\mathrm{CON}$) score quantifies the number of common out-neighbors a node
shares with the rest of the network; see \cite{CON} for more on the CON score. We let $\mathrm{CON}(u,v)$ be the number of common out-neighbors of distinct nodes $u$ and $v,$ and define
$$\text{CON}(v) = \sum_{u \in V}\text{CON}(u,v).$$

We define an $R$ value for each cycle as an aggregate measure of the cycle's performance. Before we can do that, we want to assign a value for each of the four centrality measures mentioned just above to each cycle. We do this using the community-normalized score for each measure. In the end, for each cycle $C \in \mathcal{C}$ we have a value for each of the following: spread number $r$, betweenness centrality $\mathrm{\mathrm{CNS}}(C,B)$, closeness centrality $\mathrm{\mathrm{CNS}}(C,\mathrm{CL})$, degree centrality $\mathrm{\mathrm{CNS}}(C,D)$, and CON score $\mathrm{\mathrm{CNS}}(C,\mathrm{CON})$. Each value should be further min-max normalized into the range $[0,1]$.

We introduce an aggregate measure $R$ for a cycle $C$, which incorporates the embedding information $r$ and the degree and betweenness centrality measures. Define
$$R(C) = \frac{1}{3}\left(r(C) + (1-\mathrm{\mathrm{CNS}}(C,B)) + (1-\mathrm{\mathrm{CNS}}(C,D))\right).$$
The $R$ value is intended to reflect the discussion in Section~2, which flags nodes with a high spread number but low betweenness and degree centralities as suspicious and anti-central.

\section{Results}\label{results}

Building on our previous work in \cite{prev}, we examine the cleaned Rabobank transaction network, comprising $83$ cycles (forming the collection $\mathcal{C}$) and a partition of the nodes into $11,827$ communities (forming the collection $\mathcal{P}$). The implementation we used for \textit{node2vec} is \textit{PecanPy} presented in \cite{pecanpy}. All other network algorithms were implemented using \textit{NetworkX}. All $\mathrm{CNS}$ values were used with $m = 100$ random samples, and the dimension of the embedding was fixed at $d = 32$.

\begin{figure}[htpb!]
\centering
\includegraphics[scale=0.3]{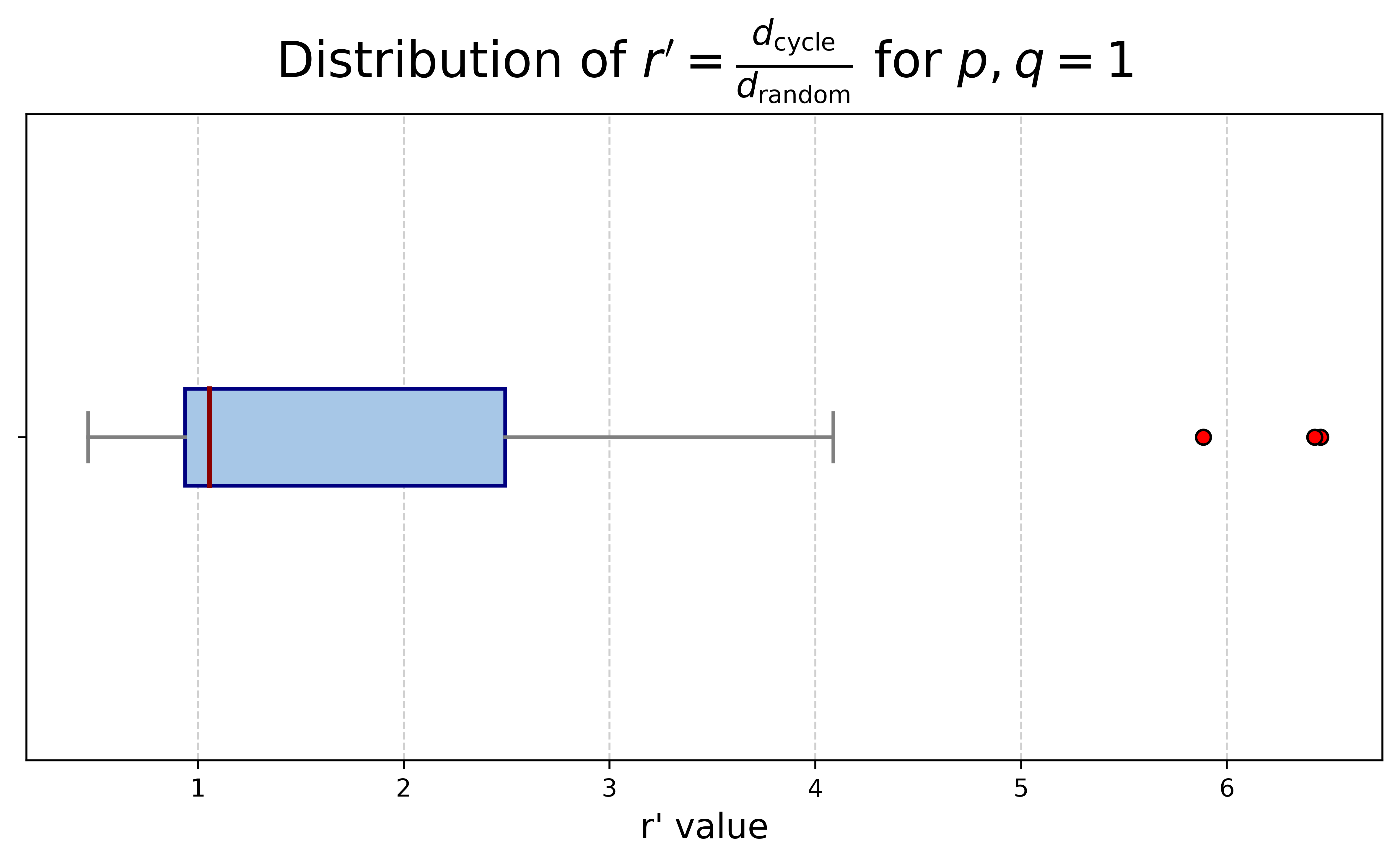}
\caption{Example distribution of $r'$ values across the 83 detected cycles when $p = q = 1$. Outliers around $r'=6$ indicate cycles that are substantially more dispersed in the embedding space compared to random subgraphs from the same community.}\label{fig:f1}
\end{figure}

Figure~\ref{fig:f1} shows the distribution of $r'$-values for $p=q=1$, a common parameter choice in the literature. Each cycle is assigned an $r'$ value, with several outliers reaching values around $6$, indicating these cycles are approximately six times more dispersed in the embedding space than random subgraphs of equal size within their community.

\begin{figure}[htpb!]
\centering
\includegraphics[scale=0.27]{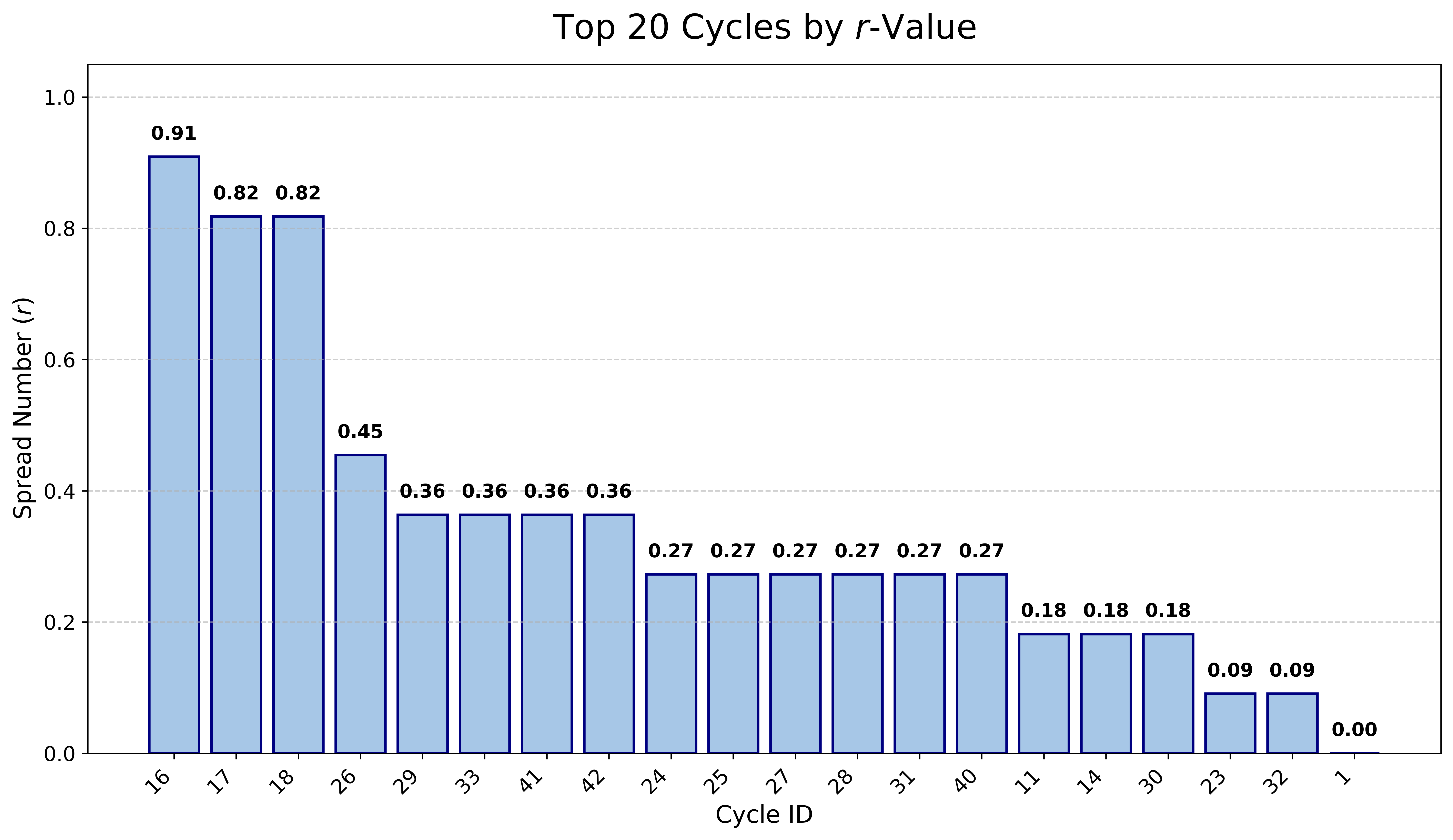}
\caption{Top-ranked cycles by spread number (or $r$). Only 19 of the 83 cycles have non-zero spread numbers, with the three highest cycles standing out sharply. }\label{fig:f2}
\end{figure}

We vary $p,q$ for robustness, and compute the spread number $r$ for each cycle. The top-scoring cycles in terms of $r$ are presented in Figure~\ref{fig:f2}. The values were computed by varying $p,q$ for $k = 8$ evenly-spaced values in $(0,2]^2$. Note that only $19$ of the original $83$ suspicious cycles have a non-zero spread number, with the top three (all belonging to the same community of size $882$) standing out sharply.

\begin{figure}[htpb!]
\centering
\includegraphics[scale=0.25]{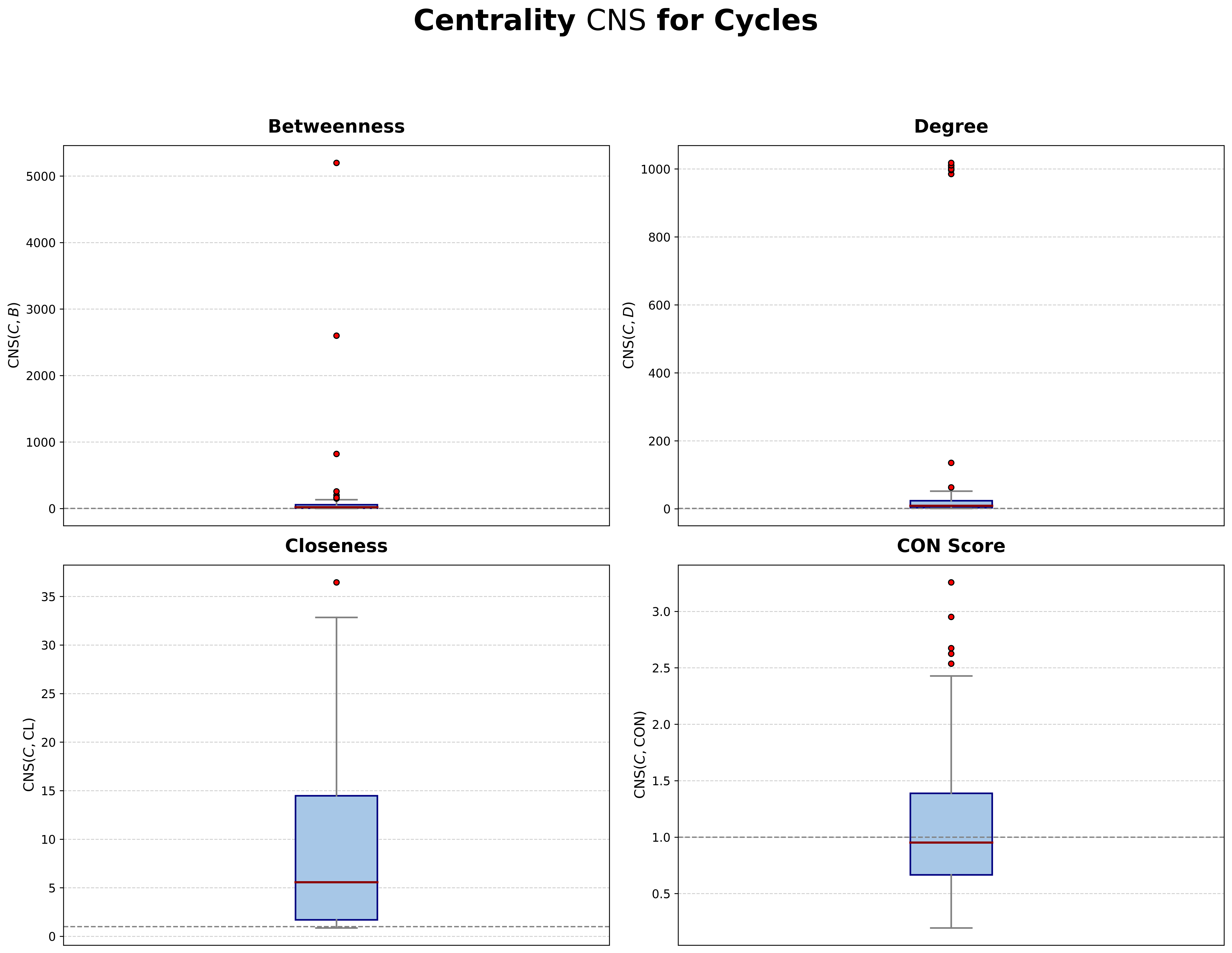}
\caption{Community-normalized centrality scores (CNS) for betweenness, degree, closeness, and CON across the 83 cycles. The dotted line marks $\mathrm{CNS}=1$, corresponding to the community baseline.}\label{fig:f3}
\end{figure}

We next examine the other measures of centrality, normalized by community, in Figure \ref{fig:f3}. We observe that there exist extreme outliers in the case of betweenness and degree centrality, but not in closeness or $\mathrm{CON}$. This supports our decision to exclude the latter pair from the aggregate score $R$. For example, we expect anti-central nodes to have low closeness in an effort to avoid detection. The CON score is a global measure that relies on synchronizing transactions among multiple accounts across the network. We therefore would not expect cycles involving money laundering activities to have nodes with a high CON score.

\begin{figure}[htpb!]
\centering
\includegraphics[scale=0.22]{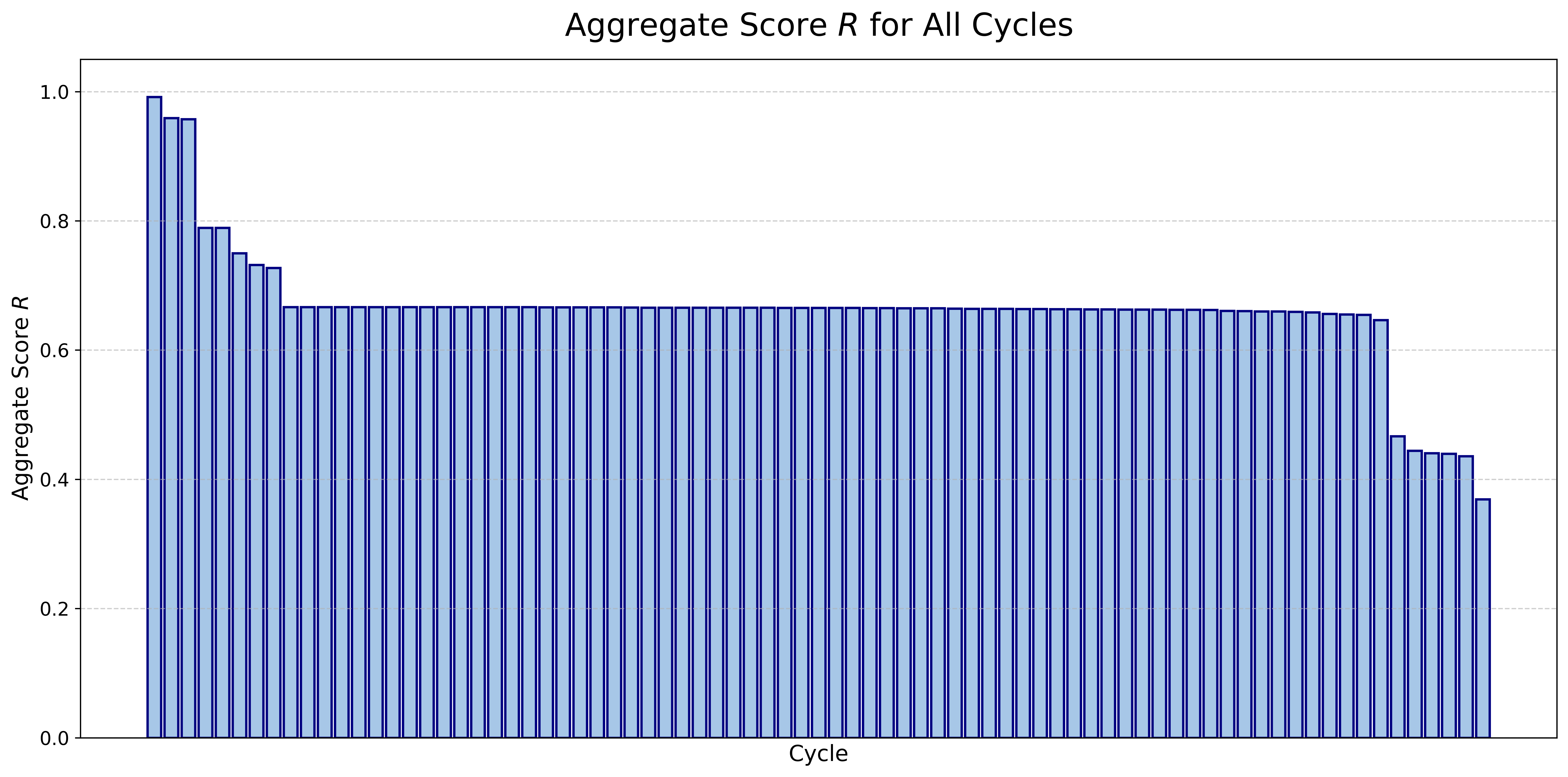}
\caption{Sorted aggregate $R$ scores across the 83 cycles. Note that a small number of cycles attain significantly higher values, identifying them as the most anti-central.}\label{fig:f4}
\end{figure}

Figure~\ref{fig:f4} displays the ranking of cycles by $R$ values. The cycles that score the highest in terms of $R$ would be considered the most anti-central and match the expected behavior of money launderers in a transaction network, as discussed in Section~3. Most cycles cluster near the mean, but a few stand out as anomalies. We note that the top three cycles in terms of spread number (see Figure~\ref{fig:f2}) are the same as those with the highest $R$ values in Figure \ref{fig:f4}.

We repeated these experiments using PageRank rather than degree in our aggregate $R$ score. However, the results regarding top-scoring cycles were identical.

\section{Discussion and Future Directions}\label{disc}

We introduced a network-embedding approach to improve anti-money laundering detection. We combined the new notion of spread number $r$, which captured how dispersed nodes in cycles are in the embedding space, with traditional centrality measures to define an aggregate score $R$. The $R$ score highlights cycles whose nodes are what we refer to as \textit{anti-central}: structurally important yet inconspicuous, reflecting the expected behavior of money launderers.

Results on the Rabobank network indicate that only a small subset of cycles achieves high $R$ values, effectively identifying suspicious nodes. If we consider only the top three scorers in terms of $R$, then we obtain three cycles comprising only seven accounts (with overlap between the three cycles), which are flagged as the most suspicious according to this methodology. Such a small number of suspicious accounts may be readily investigated by banks or regulatory agencies.

A challenge facing our method is the lack of ground-truth labels, specifically knowing in advance which accounts are involved in money laundering. Additional datasets (which are difficult to obtain given their sensitive nature) may resolve this challenge. Since money laundering activity evolves over time, analyzing datasets with more temporal network information could provide valuable insights.

Besides cycles, we may consider directed shortest paths intersecting at least one cycle. Directed paths are sequences of directed transactions that can be traced between two nodes, but which (unlike in the case of directed cycles) do not return the funds to the source. Longer shortest directed paths between nodes may also indicate an attempt to layer illicit funds.
Based on a preliminary analysis of the Rabobank banking network, after detecting directed paths of length between four and seven within communities that contained at least one cycle, we found $412,852$ paths, consisting of $7,527$ nodes, $146$ of which were in both a path and a cycle. We then examined each community and assessed the likelihood that a node is part of a directed cycle within the collection of cycles identified in previous work, assuming it is part of a directed path. Some communities were more conspicuous in terms of this measure, and we plan on continuing this approach using directed paths in future work.

Our analysis of patterns of connectivity, embedding-space dispersion, and interaction with community structure can be applied more generally to other types of complex networks beyond banking networks.
Future work could further develop the concept of anti-centrality by systematically investigating how nodes deliberately avoid traditional measures of centrality while still exerting structural influence within the network. This could be relevant to other socially determined networks.

\end{document}